\documentclass[conference]{IEEEtran}
\pdfoutput=1
\usepackage{cite}
\usepackage{graphicx}
\graphicspath{ {images/} }

\begin{document}


\title{Enhancing Engagement in Token-Curated Registries via an Inflationary Mechanism}

\author
{\IEEEauthorblockN{Yi Lucy Wang, Bhaskar Krishnamachari}
\IEEEauthorblockA{Viterbi School of Engineering\\
University of Southern California\\
Los Angeles, CA 90089\\
Email: \{wang796, bkrishna\}@usc.edu}
}
\maketitle


\begin{abstract}
Token Curated Registries (TCR) are decentralized recommendation systems that can be implemented using Blockchain smart contracts. They allow participants to vote for or against adding items to a list through a process that involves staking tokens intrinsic to the registry, with winners receiving the staked tokens for each vote. A TCR aims to provide incentives to create a well-curated list. In this work, we consider a challenge for these systems - incentivizing token-holders to actually engage and participate in the voting process. We propose a novel token-inflation mechanism for enhancing engagement, whereby only voting participants see their token supply increased by a pre-defined multiple after each round of voting. To evaluate this proposal, we propose a simple 4-class model of voters that captures all possible combinations of two key dimenions: whether they are engaged  (likely to vote at all for a given item) or disengaged, and whether they are informed (likely to vote in a way that increases the quality of the list) or uninformed, and a simple metric to evaluate the quality of the list as a function of the vote outcomes. We conduct simulations using this model of voters and show that implementing token-inflation results in greater wealth accumulation for engaged voters. In particular, when the number of informed voters is sufficiently high, our simulations show that voters that are both informed and engaged see the greatest benefits from participating in the registry when our proposed token-inflation mechanism is employed. We further validate this finding using a simplified mathematical analysis. 
\end{abstract}

\begin {IEEEkeywords}
Token-curated registry; decentralized recommendations; Blockchain
\end{IEEEkeywords}

\section{Introduction}
\label{intro}

Token Curated Registries (TCR) have been proposed as a decentralized alternative to centralized recommendation systems that can be implemented using Blockchain smart contracts~\cite{goldin2017token}. They allow participants to vote for or against adding items to the list through a process that involves staking tokens intrinsic to the registry, with winners receiving the staked tokens. A TCR aims to provide incentives to create a well-curated list, by making such a list valuable for token-holders. A TCR could be created for any topic where a set of recommendations could be useful, for example a list of universities, or restaurants, or reliable sources of data. 

One example of a real-world TCR project is the adChain registry~\cite{adChain} by MetaX, which curates a list of possible websites that could be used for digital advertising. There are now several application and development projects related to TCR such as the Ocean Protocol~\cite{ocean} and Decentralized Data Marketplace (DDM)~\cite{DDM}, which advocate using TCR to curate data sources, and Dirt Protocol~\cite{dirt} which seeks to develop tools for deploying TCR's for different domains. A recent study explored the TCR from the perspective of game theory, to understand the types of equilibria that may be possible, and concluded that good and bad voting outcomes could both be valid equilibria in some cases, resulting in problems of coordination~\cite{asgaonkar2018token}. In this work, we focus on a different aspect of TCR's: how to increase engagement by token holders to encourage them to vote. In the absence of explicit incentives to vote, it is possible for some (or even many) token holders to act as ``free-riders"  by letting others do the work of voting to curate the TCR list. 

We introduce and explore an enhancement to TCR's that is aimed to enhance engagement. In a nutshell, our proposal is to supply additional tokens (at a fixed multiplicative rate) to token holders each time that they vote.  This results in inflation since the total number of tokens is increasing, each token is worth a smaller fraction of the value of the TCR. However, participants that vote regularly (i.e. are engaged) are better able to keep up with inflation as their token supply increases at the rate of inflation, compared to participants that are disengaged. In a TCR where a majority of the participants make good decisions when they vote, the intrinsic TCR value will naturally increase as the quality of the TCR improves, and with our proposed enhancement, we can hypothesize that over time a significant fraction of the ``wealth" inherent in the TCR economy would accumulated by those voters that are both engaged enough to vote often and informed enough to vote correctly. This in turn creates an incentive for more engaged and informed voters to become participants of a given TCR. 

In order to evaluate this hypothesis, we build a quantitative model that is purposely simplified to yield 
useful insights. Our modeling has two major components: a) a probabilistic voter behavior model, and b) a metric for TCR quality.  For voter behavior, we assume that each voter is engaged with certain probability (else disengaged), and independently of that, is informed with a certain probability (else uninformed). A voter that is engaged votes with a higher probability than a voter that is disengaged. And a voter that is informed has a higher defined probability of making correct decisions when voting, compared to a voter that is uninformed. For measuring TCR quality, we focus on TCRs where for all items being considered, their suitability for the list can be described in a binary manner - i.e. each item is inherently either good or bad.  We adopt a very simple metric that we refer to as linear unit reward and penalty (LURP) metric, which adds a unit to each "correct" decision (to add a good item to the registry or reject a bad item) and subtracts a unit for each "incorrect" decision (to reject a good item or add a bad item). Although in general the total monetary value (or  capitalization) associated with the TCR could be modeled as any monotonic function of this quality metric. For ease of exposition, in this work, we simply use the identity function, i.e., the total value of the TCR at a given time is assumed to be the same as its LURP metric at that time.

We conduct simulation-based experiments using our model to evaluate our hypothesis that the proposed inflation mechanism results in greater wealth for informed and engaged voters (assuming a majority of voters are informed) compared to the control of not using inflation. The simulation tool we have developed may be of independent interest to the community and we have therefore made it publicly available as an open source tool online at \url{https://github.com/ANRGUSC/TCRsim}. We also present a mathematical analysis of a further simplified setting that validates the key findings from the simulations. 

In the following section, we describe our TCR model in more detail. After that, we present our simulation parameters and results, followed by mathematical analysis of a simplified setting. Finally, we present a summary of our contribution and discuss possible caveats and future work. 

\section{A TCR Model with Token Inflation}
\label{protocol}

\subsection{Basic TCR model}
We model the TCR very simply as follows:
\begin{itemize}
\item There are $N$ number of voters, that vote successively on each of $L$ number of items.
\item An item is voted into the registry if more than $50\%$ of the voters vote to add it in. 
\item At each round of voting for a given new item, all voters are required to stake a certain amount of tokens.
If the voter sides with the majority vote in that round, she/he gets the total stake divided by number of winning voters; if a voter sides with the minority vote in that round, she/he will lose all staked tokens. If the voter does not have the required number of tokens to be staked in a given round she/he cannot vote.
\item We assume that we are dealing with an objective TCR, where each item is objectively either "good" or "bad" - and that a voter that votes correctly would vote to add a "good" item to the registry and to not add a "bad" item to the registry (the opposite being true if the voter votes incorrectly). 
\end{itemize}

Some notes are in order. In traditional TCR, the stake pool also includes tokens staked by whoever proposes a new item to be added to the registry - we choose to omit this in our modeling for simplicity; assuming a sufficiently large pool of voters, it should have negligible impact on the outcomes. Another particular decision we make is to have all of the staked tokens of the minority voters transferred to the majority voters. 

\subsection{Proposed inflation mechanism}
In traditional TCRs the minimum amount that must be staked by each voter is kept constant. With the aim of 
incentivizing engagement, we propose two key changes:
\begin{enumerate}
    \item After each round of voting, we propose to increase the tokens available to those participants that have voted by a given multiplicative factor. Specifically,  we introduce an inflation rate parameter $\delta$. Each voter $j$ that votes in a round $t$ gets its tokens $T_j$ inflated as follows: $T_j(t+1) = T_j(t)*(1+ \delta)$, where $T_j(t)$ and $T_j(t+1)$ represent that voter's number of tokens before (at the end of some round t) and after inflation (before voting on item t+1). Voters that do not vote see no change in the number of tokens they hold. Note that setting $\delta=0$ recovers the original, un-inflated TCR.
    \item We increase the minimum amount of tokens that should be staked over time to keep pace with the growing number of tokens in the system. Specifically, say that before voting for the first item, each voter has a certain number of initial tokens ($T(0)$), and must stake a certain number of tokens $S(0)$ that is assumed to be a fixed percentage of the initial tokens  before voting.  In subsequent rounds of voting the required stake amount $S(t)$ is increased to keep pace with the inflation as follows, where $T_{total}(t)$ represents the total number of tokens: $S(t) = \frac{S(0)}{T_{0}}\times \frac{T_{total}(t)}{N}$
\end{enumerate}

\subsection{Voter behavior model} For the purpose of simulation, we further model voters as being classified into four categories based on two orthogonal binary classifications: they are either informed or uninformed; engaged or disengaged. We use two i.i.d. Bernoulli distributions for these classifications to determine which category each voter belongs to. We describe below additional parameters used in the simulations:

\begin{itemize}
\item $p_{E}$ is the probability a given voter is engaged, and hence probability $1-p_{E}$ is the probability the voter is disengaged. If a voter is engaged, she/he participates in the vote for a given item with Bernoulli probability $p_{VE}$, and if a voter is disengaged, she/he participates in the vote for a given item with probability $p_{VD}$ which is assumed to be smaller than $p_{VE}$. In this work, we have made a simplifying assumption that the engagement characteristic of a voter does not change over time.  
\item $p_{I}$ is the probability a given voter is informed, and hence $1-p_{I}$ is the probability that the voter is uninformed.   If a voter is informed, if voting, she/he votes correctly with probability $p_{VCI}$, and if a voter is uninformed, if voting, she/he votes correctly with probability $p_{VCU}$ 
\end{itemize}

\subsection{TCR quality model}

For our evaluation, we further need some measure of the quality of the TCR, and how that in turn translates to value. We use a very simple linear model that we refer to as the linear unit reward and penalty (LURP) metric, which assumes that each correct or valid vote (to add a good item or keep out a bad item) increases the quality of the TCR by a unit amount, while each incorrect or invalid vote (to leave out a good item or add in a bad item) decreases the quality by a unit amount. In general the value of the TCR could be expected to be monotonic with the quality metric. For simplicity, we assume the identity function so that we describe the inherent value of the TCR (the total value of all tokens in the economy) at any time by the LURP metric of the TCR. Thus, we can write the total value of the TCR at any round t, $W_{tot}(t)$ in terms of the number of corect and incorrect votes upto that round (denoted $V_{correct}$ and $V_{incorrect}$ respectively): 
\begin{equation}
    W_{tot} = V_{correct} - V_{incorrect}
\end{equation}

It should be noted that in reality the LURP metric is only computable if the objective/ground truth about each item is known, and in practice the total value of the TCR may not be simply equal to the LURP metric, but these are useful modeling assumptions that allow us to evaluate the proposed mechanism.

For voters in each of the four classes of voters (informed-engaged, informed-disengaged, uninformed-engaged, and uninformed-disengaged), we can measure their average wealth in a given class $A$, denoted as $\hat{W}_A$ as follows: 
\begin{equation}
    \hat{W}_A = \frac{W_{tot}}{T_{tot}} \times \frac{T_A}{n_A}
\end{equation}

where $W_{tot}$ is the LURP-based total value of the TCR, $T_{tot}$ is the total number of tokens, $T_A$ is the total number of tokens among voters in class $A$, and $n_A$ is the total number of voters in class $A$.

In the next section we show through simulations that with inflation, engaged voters 
generally have greater wealth than without inflation, while the quality or value of the TCR depends crucially 
on the fraction of voters that are informed. This creates an incentive for the TCR to attract more engaged and informed voters, which, in turn, should further improve the quality of the TCR. 

\section{Simulation and Results}
\label{demoResults}
   

We created a simulator using Python called TCRsim (available online at 
https://github.com/ANRGUSC/TCRsim)
that allows us to evaluate and understand the impact of token inflation. 
Table~\ref{table:params} shows the parameters we used in our simulations. 
In the following we present a set of illustrative results from these simulations.
In these simulations we essentially vary two of the parameters: 
a) the probability of being informed (we vary this to take on three
values: 0.1, 0.5, and 0.9) and b) whether or not inflation is used (with a 
default value of $2\%$). 

\begin{table}[ht]
\centering
\begin{tabular}{|l|l|l|}
	\hline
	Parameter & Values \\
	\hline
    $L$ & 50 \\
	\hline
    $N$ & 100 \\
	\hline
     $T(0)$ & 100  \\
	\hline
    $S(0)$ & 5.0 \\
	\hline
    $\delta$ & 0.0, 0.2 \\
 	\hline
	$p_E$ & 0.5 \\
	\hline
    $p_I$ & 0.1, 0.5, 0.9 \\
	\hline
    $p_{VE}$ & 0.8 \\
	\hline
    $p_{VD}$ & 0.2 \\
	\hline
    $p_{VCI}$ & 0.85 \\
	\hline
    $p_{VCU}$ & 0.15 \\
	\hline
\end{tabular}

\caption{Simulation Parameters}
\label{table:params}
\end{table}

\begin{figure*}[p]
   \centering
   \includegraphics[scale=.4]{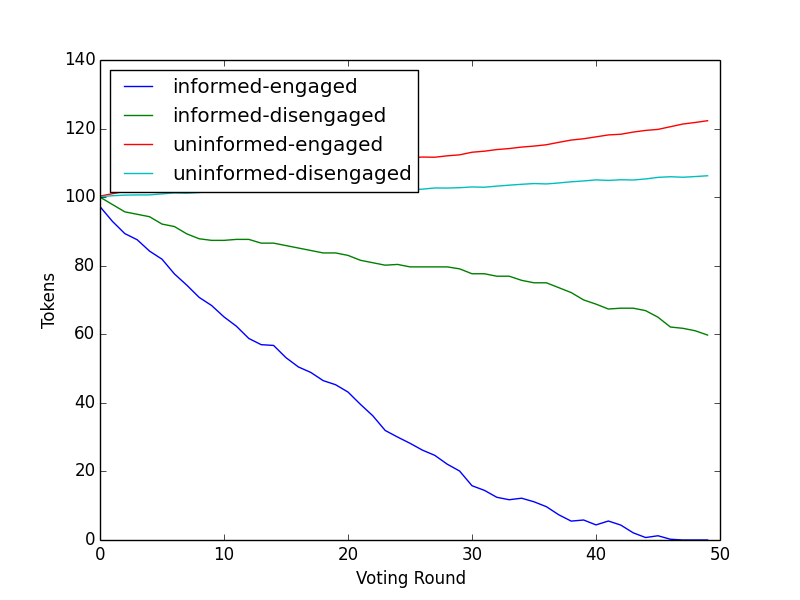}
   \includegraphics[scale=.4]{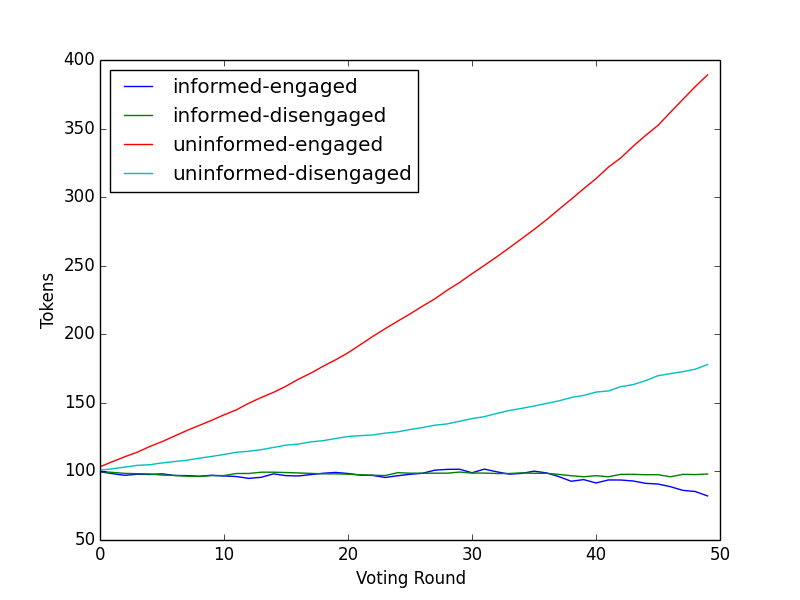}
   \includegraphics[scale=.4]{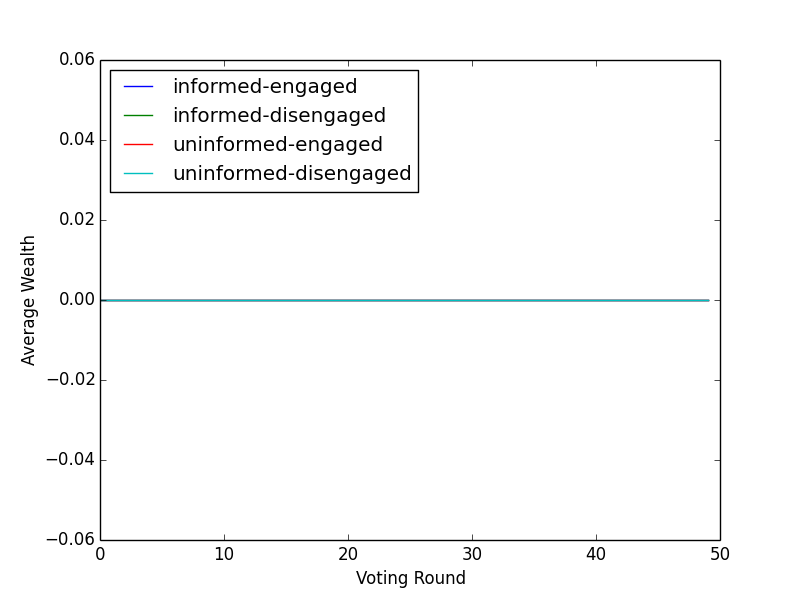}
    \includegraphics[scale=.4]{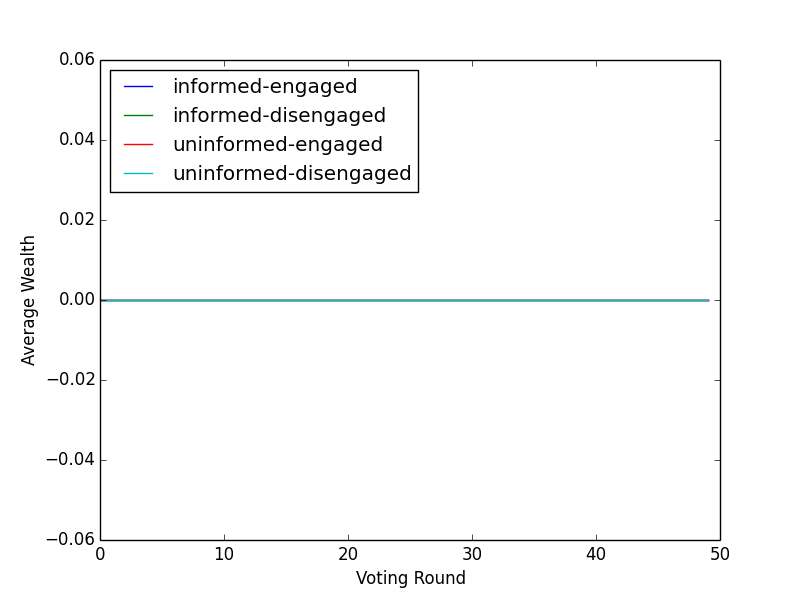}
   \includegraphics[scale=.4]{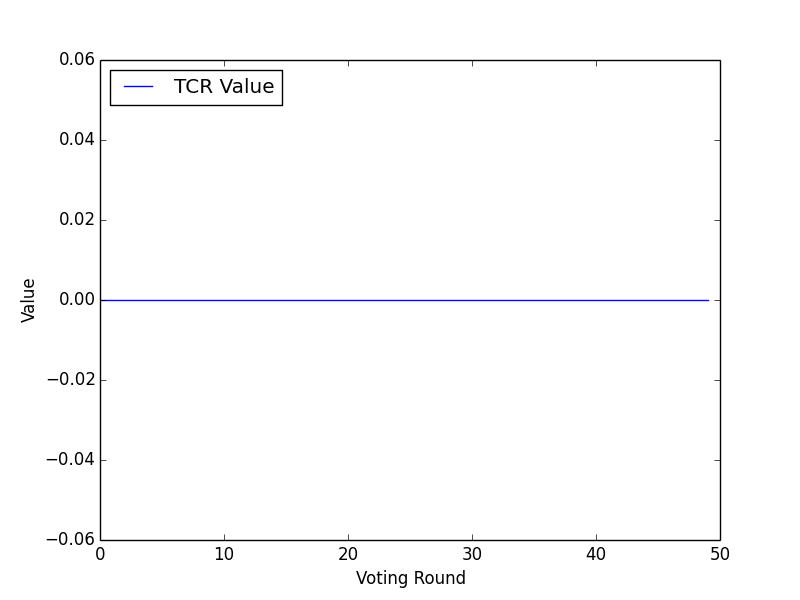}
   \includegraphics[scale=.4]{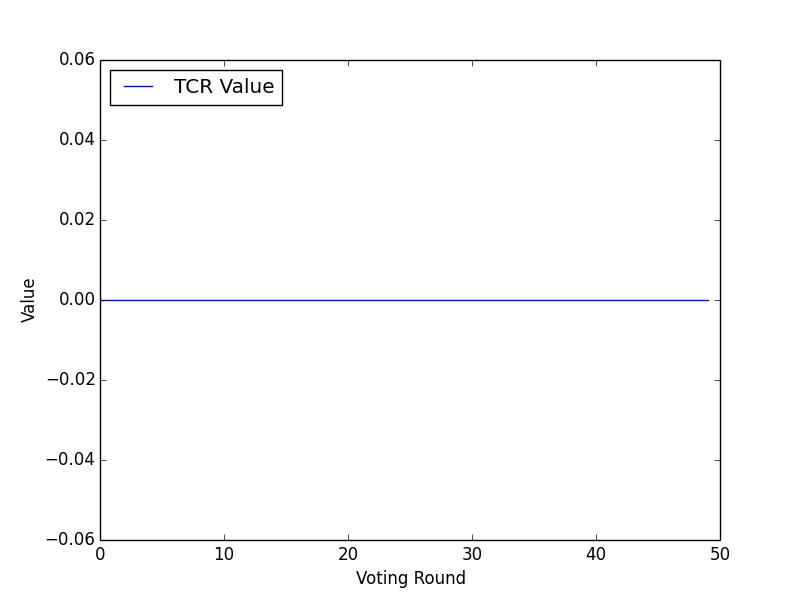}
    \caption{Simulation results for $p_I = 0.1$, left column: without inflation; right column: with 2$\%$ inflation}
    \label{fig:pi1}
\end{figure*}

\begin{figure*}[p]
   \centering
   \includegraphics[scale=.4]{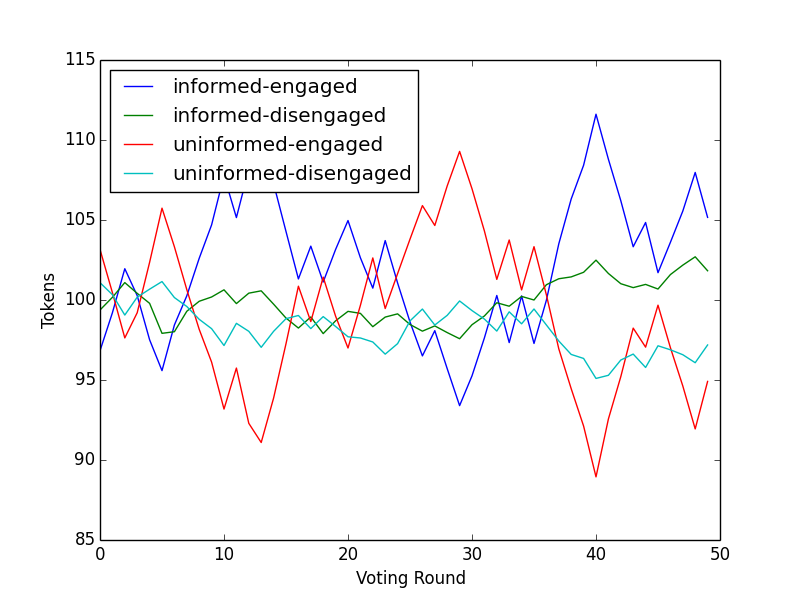}
   \includegraphics[scale=.4]{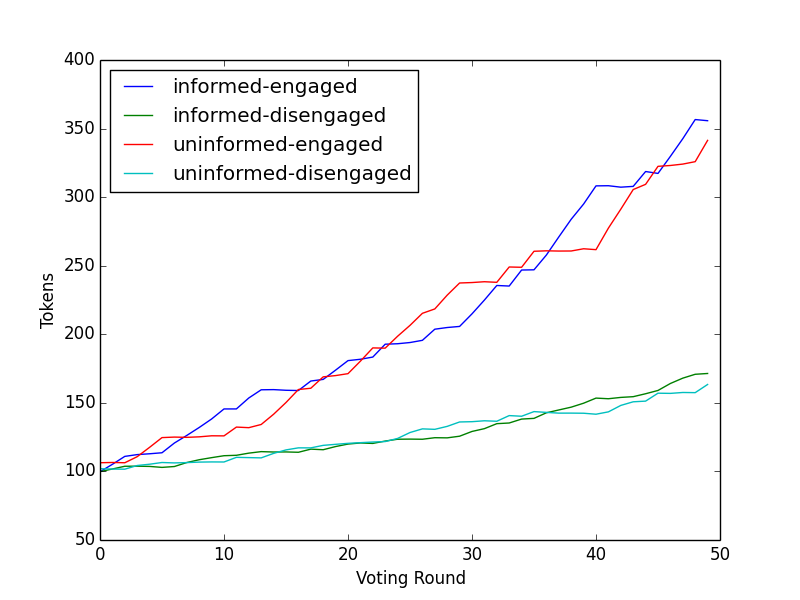}
   \includegraphics[scale=.4]{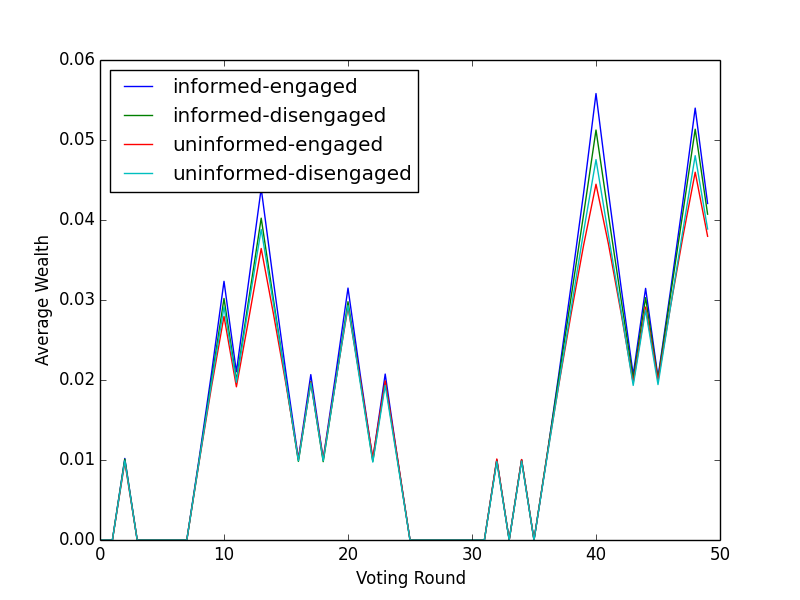}
    \includegraphics[scale=.4]{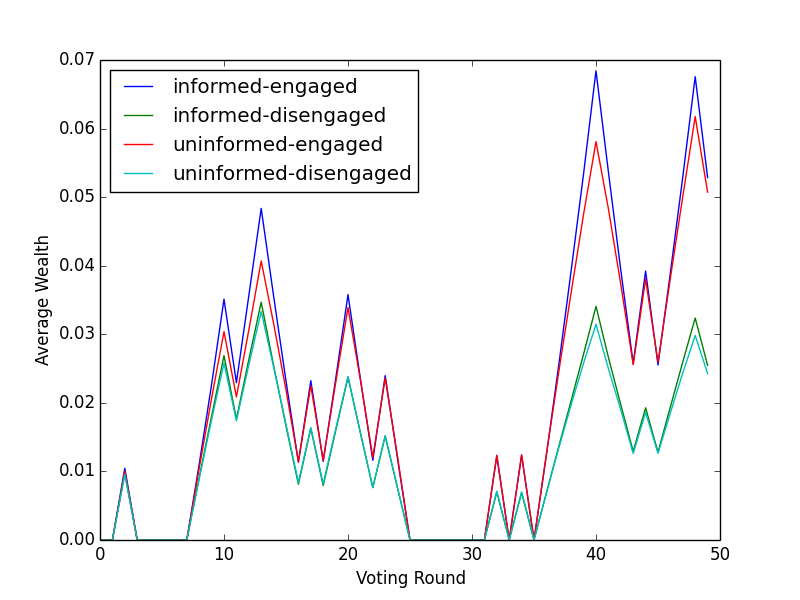}
   \includegraphics[scale=.4]{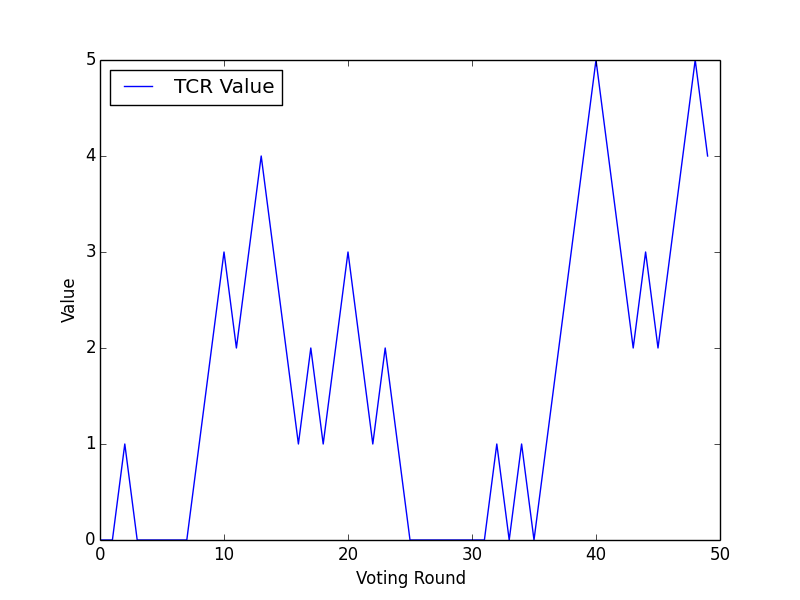}
   \includegraphics[scale=.4]{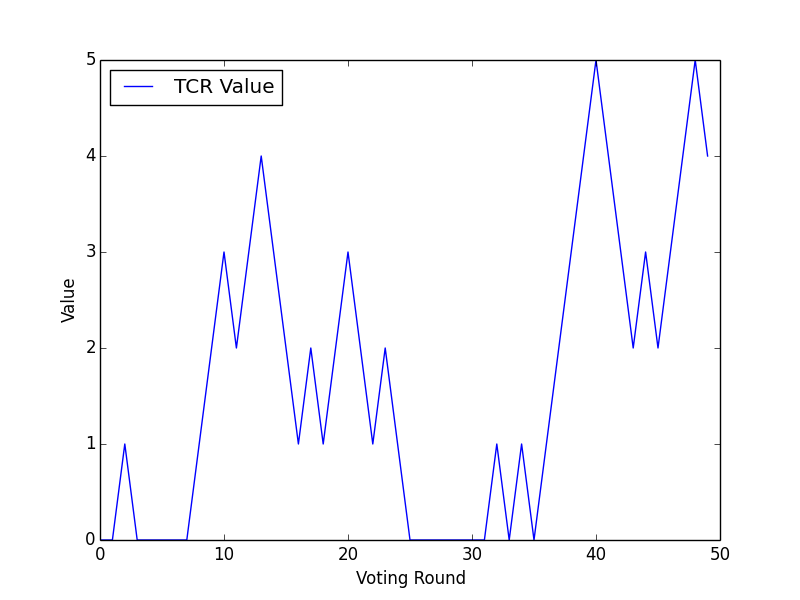}
    \caption{Simulation results for $p_I = 0.5$, left column: without inflation; right column: with 2$\%$ inflation}
    \label{fig:pi2}
\end{figure*}

\begin{figure*}[p]
   \centering
   \includegraphics[scale=.4]{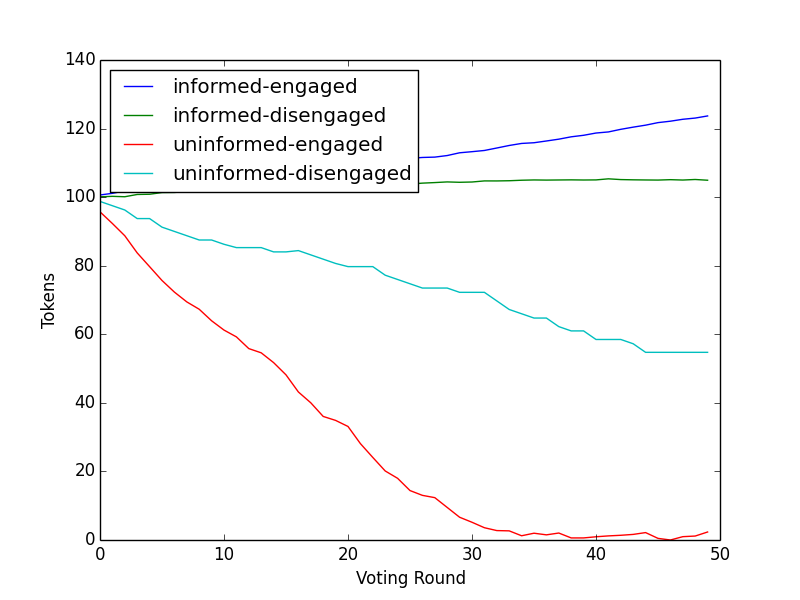}
   \includegraphics[scale=.4]{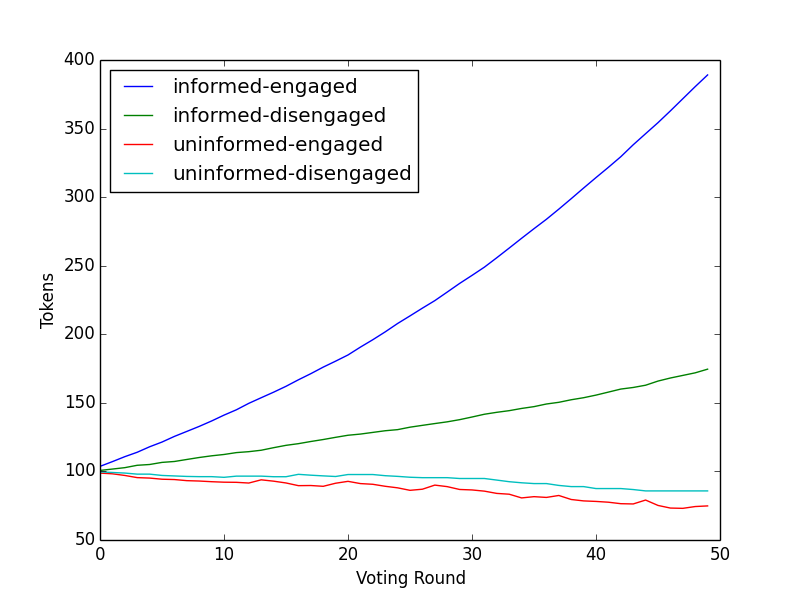}
   \includegraphics[scale=.4]{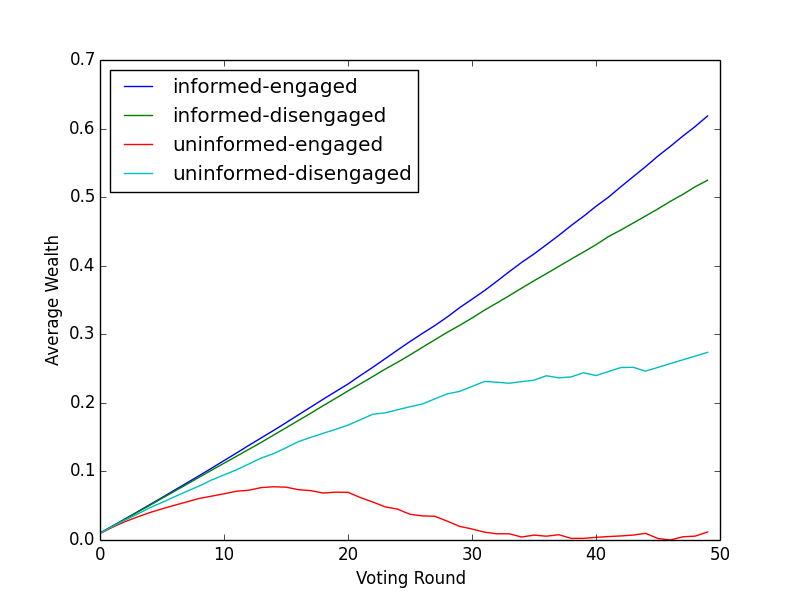}
    \includegraphics[scale=.4]{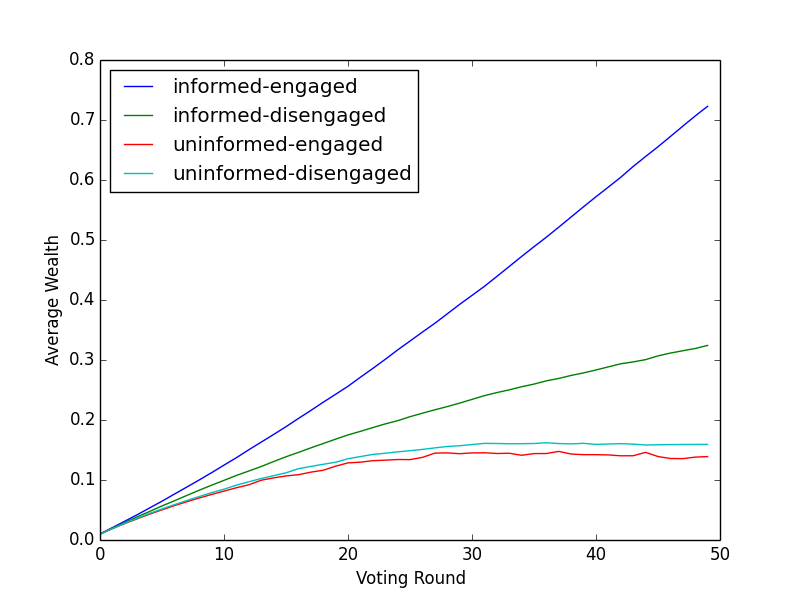}
   \includegraphics[scale=.4]{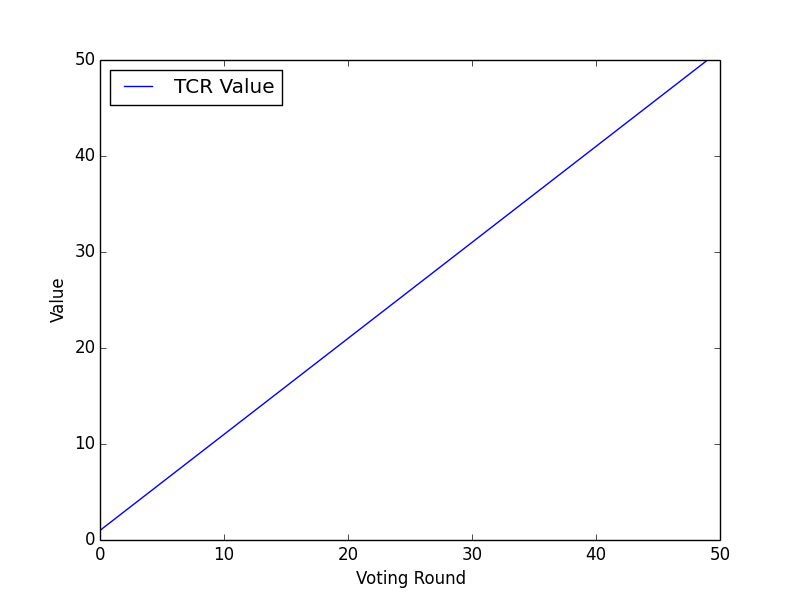}
   \includegraphics[scale=.4]{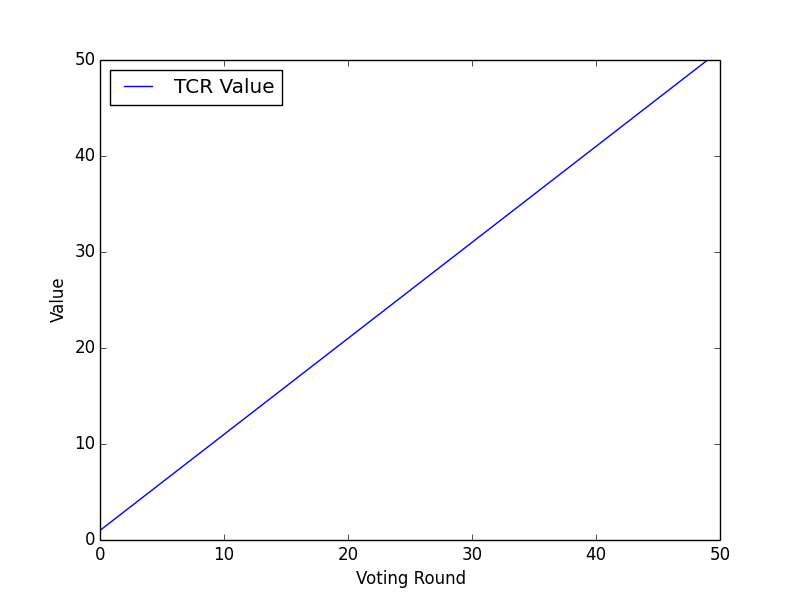}
    \caption{Simulation results for $p_I = 0.9$, left column: without inflation; right column: with 2$\%$ inflation}
    \label{fig:pi3}
\end{figure*}

Figures~\ref{fig:pi1},~\ref{fig:pi2}, and~\ref{fig:pi3} present the results of our simulations. 

The following are the key findings from the simulations:
\begin{itemize}
    \item When the probability of voters being informed, and hence voting correctly on items, is low (a bad case for the TCR), as considered in figure~\ref{fig:pi1} then the TCR value and hence the wealth of voters in each class, stays at zero. This is because most votes are influenced strongly by the uninformed voters and result in a poor quality TCR with virtually no valid or correctly added items. The uninformed voters have a high number of tokens with or without inflation but it is immaterial given the zero-value of the TCR. Such a TCR in practice will struggle to survive, regardless of whether inflation is used. 
    \item In the case where the probability of being informed is moderate (0.5), simulations show high variance; here we present one particuar run. We see that the value of the TCR fluctuates, and so does the fortune of the various classes of users. We can see that because of the even probability of correct votes, all classes of voters have about the same wealth in case of the traditional TCR. However, with inflation, we can see that the engaged voters (both informed and uninformed) have higher wealth on average than disengaged voters
    \item When the probability of being informed is high (a good case for the TCR), as considered in figure~\ref{fig:pi3}, then we see that TCR value increases linearly with each vote (since all votes end up being correct and valued). In either case, informed voters get more tokens and wealth compared to uninformed voters; we can see that with inflation, the engaged informed voters have significantly higher wealth compared to unengaged voters, thus giving an incentive to voters for being more engaged, with informed voters benefiting the most because they also win more of the staked tokens across the voting process.
\end{itemize}

\begin{figure}[t]
   \centering
   \includegraphics[scale=.4]{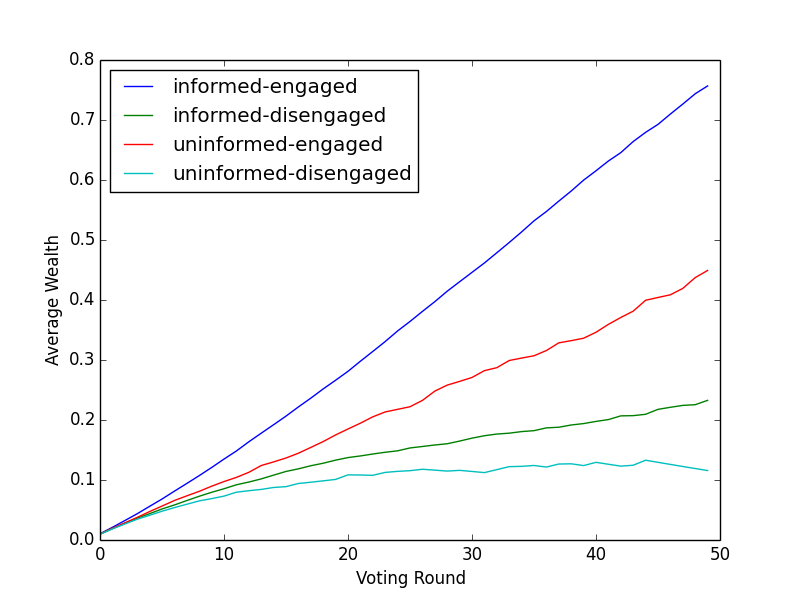}
    \caption{Results for $p_I = 0.9$ with an inflation rate of 5\%}
    \label{fig:inf5}
\end{figure}

Finally, to evaluate the impact of increasing the inflation rate further, we conduct one additional simulation for the case where $p_I$ is 0.9 but where the inflation rate is increased to $5\%$. The results of this experiment in terms of the average wealth for users of each class is shown in figure~\ref{fig:inf5}. We see that compared to the case of $2\%$ inflation for the same high probability of informed voters (seen in the right column of~\ref{fig:pi3}), this case results in even higher wealth for uinformed but engaged voters compared to informed disengaged voters. This shows that, as expected, higher inflation gives even higher incentive to voters for being engaged, to the point that even uninformed voters can have a significant benefit. From a realistic perspective however, in practice, this greater incentive for engagement has to be weighted against the concern that engaging uninformed voters too much runs the risk of having a lower quality TCR, thus a low/moderate inflation rate might be preferred to having a high inflation rate. 


\section{Mathematical Analysis}
\label{analysis}

To gain more insight, we consider and analyze 
a simple, idealized setting that is amenable to 
closed-form modeling. For the analysis in this section,
we assume that:
\begin{itemize}
    \item Disengaged participants never vote, while engaged participants always vote
    \item Informed voters are always correct while uninformed voters are always incorrect
    \item The number of informed engaged voters $n_{IE}$ is more than the number of uninformed engaged voters $n_{UE}$ so that correct decisions are made for each item. (Also, we denote by $n_{ID}$ and $n_{UD}$ the number of informed disengaged and uninformed disengaged voters, respectively).
    \item  Each time there is a vote, each participant is required to stake an amount that is equal to $\sigma$ fraction of the total token held by each uninformed participant. 
\end{itemize}

Under these assumptions, the number of tokens held by disengaged participants 
after voting on the $k^{th}$ item remains constant: 
\begin{equation}
    T_{ID}^{(k)} = T_{UD}^{(k)} = T_0
\end{equation}

The number of tokens held by engaged, but uninformed participants varies as follows (they lose the 
staked amount which is a $\sigma$ fraction of their holdings at each round and experience the 
token inflation of $1+\delta$ for voting in each round :
\begin{equation}
    T_{UE}^{(k)} = T_0 (1 - \sigma)^k (1 + \delta)^k 
\end{equation}

Note that so long as $(1 + \delta) > \frac{1}{1 - \sigma}$ the uninformed engaged participants will see their
token holdings increase, if the two terms are equal then their token holdings will remain constant, and if the inequality is reversed the uninformed engaged participants will see their token holdings decrease. 

The number of tokens held by engaged but informed participants shows an increase due to the stake winnings
of each round as well as due to the inflation for voting. 
\begin{eqnarray}
    T_{IE}^{(k)} &=& ( T_{IE}^{(k-1)} + ( T_{UE}^{(k-1)} \sigma \frac{n_{UE}}{n_{IE}} ) ) (1+\delta) \\
              &=& (T_{IE}^{(k-1)} + T_0 \sigma \frac{n_{UE}}{n_{IE}} (1 - \sigma)^{(k-1)}  (1 + \nonumber \\ & & \delta)^{(k-1)}) (1+\delta)
\end{eqnarray}
By solving the recursive expression, the above can be simplified to:
\begin{eqnarray}
   T_{IE}^{(k)} &=&  T_0 (1+\delta)^k (1 + \sigma \frac{n_{UE}}{n_{IE}} \sum_{n=0}^{k-1} (1 - \sigma)^n) \\
             &\approx& T_0 (1+\delta)^k ( 1 + \frac{n_{UE}}{n_{IE}})
\end{eqnarray}
, where the approximation holds for sufficiently large number of iterations.

From this it can be seen that the number of tokens held by 
informed engaged voters in this cases always increases exponentially, 
compounded by the inflation rate. 

The total number of tokens in the TCR after $k$ votes is given as
\begin{eqnarray}
    T_{tot}^{(k)} & = &  T_0 (n_{ID} + n_{UD} + n_{UE}(1-\sigma)^k(1+\delta)^k + \nonumber \\  & & n_{IE}(1+\delta)^k(1+\frac{n_{UE}}{n_{IE}})) 
\end{eqnarray}

If we model the total value of the TCR again as the number of correct decisions, 
the value per token after $k$ votes becomes:
\begin{equation} 
\frac{W_{tot}^{(k)}}{T_{tot}^{(k)}} = \frac{k}{T_{tot}^{(k)}}
\end{equation}

Then the value per token varies over time as $O(\frac{k}{(1+\delta)^k})$. Taking into account the number of tokens held by each class of voters as analyzed above and observing their functional form, it follows that: 
\begin{itemize}
    \item The wealth of engaged informed participants increases linearly.
    \item The wealth of engaged uninformed participants may initially increase sub-linearly (if the inflation rate is sufficiently high in relation to the stake percentage) but will eventually decrease to 0 if enough votes are held. 
    \item The wealth of unengaged participants may also increase initially for a few votes but will also soon decrease and tend to 0. 
\end{itemize}

Thus this simplified analysis also reinforces the point made through the more detailed simulations, that the proposed inflation mechanism gives the most incentive to engaged and informed voters (so long as the number of informed engaged voters is higher than the uninformed engaged voters). 

\section{Conclusion and Future Work}
\label {conclusion}

The contributions of this work are as follows. 
\begin{itemize}
\item We have proposed in this paper a mechanism to 
incentivize engagement using token inflation, specifically, 
increasing the number of tokens for users that participate
in votes. 
\item We have presented a simple (albeit static) model of 
voter engagement and informed-ness along with a simple model 
of TCR value that allows us to evaluate
the impact of the token-inflation mechanism. 
\item We wrote a custom simulator in Python called TCRSim that we make
publicly available, which we have used to evaluate the proposed mechanism. 
\item Our simulation results show that i) the percentage of informed voters
has a significant impact on the TCR value, and ii) with the inflation 
mechanism, engaged voters end up with greater wealth (corrected for inflation) 
than in the traditional TCR without token-inflation, while unengaged voters 
have less wealth. This shows that the proposed modification provides an incentive
for voters to be more engaged. 
\end{itemize}

Though we did not explicitly address how the token inflation mechanism would be implemented
in a particular TCR, it is quite feasible to implement it; one possibility would be 
through a smart contract that automatically tracks and rewards the voters with new tokens 
minted after each round of voting. Another approach suitable for TCR's with a maximum number of 
candidates to be considered would be to pre-mint and set aside a number of tokens for engagement
incentives to be dispensed to voters after each round. 

In future work, we would like to consider how the incentives could potentially affect the 
distribution of engaged/engaged voters in the system over time, for example, by considering 
the decision to vote or note as strategies from a game theoretic perspective, and how that 
in turn could impact TCR quality and the distribution of wealth. We could also consider more
complex quality metrics and ways in which the quality of votes in the TCR relate to its value. 

Finally, a caveat is in order. A crucial assumption in our work has been that a 
TCR whose total number of tokens increases exponentially but somewhat
predictably over time (a form of controlled inflation) will be acceptable in the
real world. Essentially anyone that holds a fixed amount of tokens without participating in the voting
will see their value decrease steadily due to inflation, so that the tokens will be
preferred to be held and used only by those actively involved with voting on the TCR. 
This means that the token by itself has little value outside the TCR-economy for passive
investor; However, a carefully-designed derivative that bets on the future of the TCR value could be 
of interest to such a passive investor assuming the percentage of informed voters is 
sufficiently high to build that value. How this plays out in real systems might be best 
explored empirically through projects that implement such a mechanism. 

\bibliographystyle{IEEEtran}
\bibliography{main}

\end{document}